# FACTORS INFLUENCING QUALITY OF MOBILE APPS: ROLE OF MOBILE APP DEVELOPMENT LIFE CYCLE

Venkata N Inukollu[1], Divya D Keshamoni[2], Taeghyun Kang [3] and Manikanta Inukollu[4]

[1]Department of Computer Science Engineering, Texas Tech University, USA
[2] Rawls College of Business, Texas Tech University, USA
[3] Department of Computer Science Engineering, Wake forest university, USA
[4]Department of Computer Science, Bhaskar Engineering College, India

### ABSTRACT

*In this paper, The mobile application field has been receiving astronomical attention from the past few years due to the growing number of mobile app downloads and withal due to the revenues being engendered .With the surge in the number of apps, the number of lamentable apps/failing apps has withal been growing.Interesting mobile app statistics are included in this paper which might avail the developers understand the concerns and merits of mobile apps.The authors have made an effort to integrate all the crucial factors that cause apps to fail which include negligence by the developers, technical issues, inadequate marketing efforts, and high prospects of the users/consumers.The paper provides suggestions to eschew failure of apps. As per the various surveys, the number of lamentable/failing apps is growing enormously, primarily because mobile app developers are not adopting a standard development life cycle for the development of apps. In this paper, we have developed a mobile application with the aid of traditional software development life cycle phases (Requirements, Design, Develop, Test, and, Maintenance) and we have used UML, M-UML, and mobile application development technologies.*

### KEYWORDS

*Mobile applications, low quality/bad apps, mobile apps marketing, Mobile Application development, Mobile Software Engineering, M-UML, UML*

## 1. INTRODUCTION

"A mobile app, short for mobile application or just app, is an application software designed to run on perspicacious phones, tablet computers and other mobile devices"[1]. An App makes sense or is desired if the goal is to have an interactive engagement with users, or to provide an application that requires to work more akin to a computer program than a website [2]. Apps are available via distribution platforms on concrete app stores. There are free as well as paid apps. There are few apps which initially are available for free, but later a minimum fee is required to relish premium benefits. “The iphones’ powerful software, revolutionary user interface, and powerful development platform had driven an almost overnight explosion of apps ” [6]. Most widely used smartphones for the mobile apps are iPhone, BlackBerry, Android phone or Windows Phone. For apps with a price, generally a percentage, 20-30%, goes to the distribution provider and the rest goes to the producer of the app [1]. According to mobile stats [2], the number of apps installed by the average perspicacious phone user (Global) is 26. So this number limpidly shows that apps are the expedient through which consumers want to consume content on mobile phones.

Originally mobile apps were offered for informational and productivity purposes that included email, calendar, contacts, calculator and weather information. With the rapid magnification in the technology and users' prospects the developer implements expanded into other categories such as mobile games, GPS, banking, ticket purchases, social media, video chats, factory automation, location based services, fitness apps and recently mobile medical apps [1].

An app can extract content and information from the internet in a similar fashion to a website, but it can also download the content so that it can be utilized later in the absence of Internet connection which is a great advantage [3]. So apps that do not need internet connection can be used “anywhere and everywhere” i.e. app can be used offline. Few disadvantages of the popularity of mobile apps has perpetuated to elevate, as their use has become more and more prevalent across mobile phone users. This is pellucidly evident from the numbers given by [2] i.e. Total projected Mobile app downloads in 2013 is 102 billion and the Total projected mobile app revenue in 2013 $26 billion.There are several websites and a few articles that have captured the statistics of mobile apps in terms of the number of developers growing each year, the number of apps increasing every year, revenues generated from apps, the number of apps that are appearing on different platforms and most popularly used apps on different platforms. Not many websites/articles verbalize about the number of apps being deleted, the number of users deleting apps, the number of good apps vs bad apples, factors that tempt users to delete apps, the elements that cause bad apps. There is limited literature on how good apps can be made great apps and how bad apps can be amended to become good and great.

There have been some seminal studies on how to improve apps [6]. In this paper, we add to this work by presenting the role of different elements (developers, users, technical details,) in the making of bad apps. The paper proceeds by giving the current statistics in the mobile apps field and then perpetuates with factors that cause lamentable apps, followed by suggestions to surmount those elements.

## 2. MOBILE APP MARKET STATISTICS

According to the “World Mobile Applications Market - Advanced Technologies, Global Forecast (2010-2015)” [1] there were about 6.4 billion (free, paid, and ad supported) applications, that were downloaded in 2009 alone which generated revenues of $4.5 billion in the same year. Apple ruled this market with 2.5 billion downloads from its store in 2009.Later, other market players like Android, Google, and Nokia have started creating a marketplace for themselves in the mobile apps field with the emerging smart phone market.

The market research conducted by IDC [4,5] predicts that the market for mobile applications will continue to accelerate as the number of downloaded apps is expected to increase from 10.9 billion worldwide in 2010 to 76.9 billion in 2014. Similar growth will be observed in the revenues of mobile apps(worldwide), surpassing $35 billion in 2014.

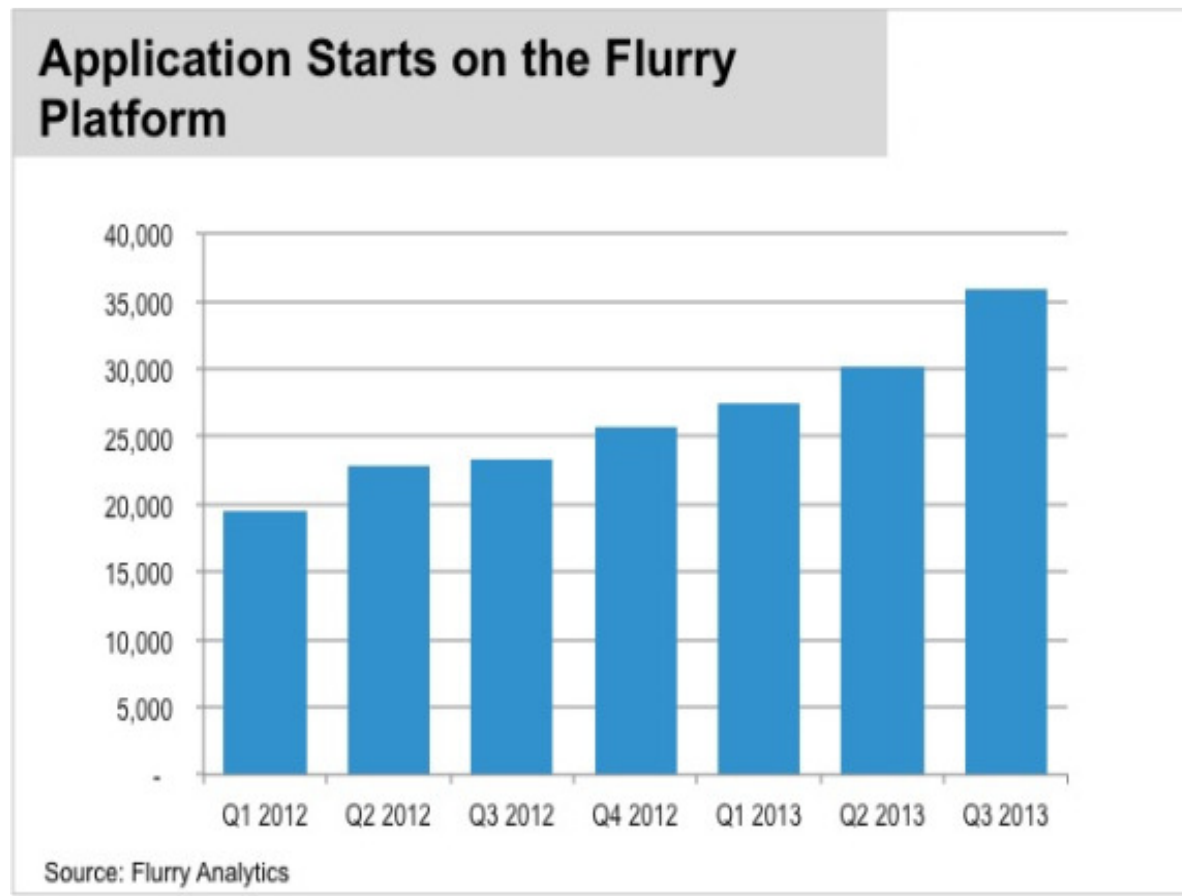

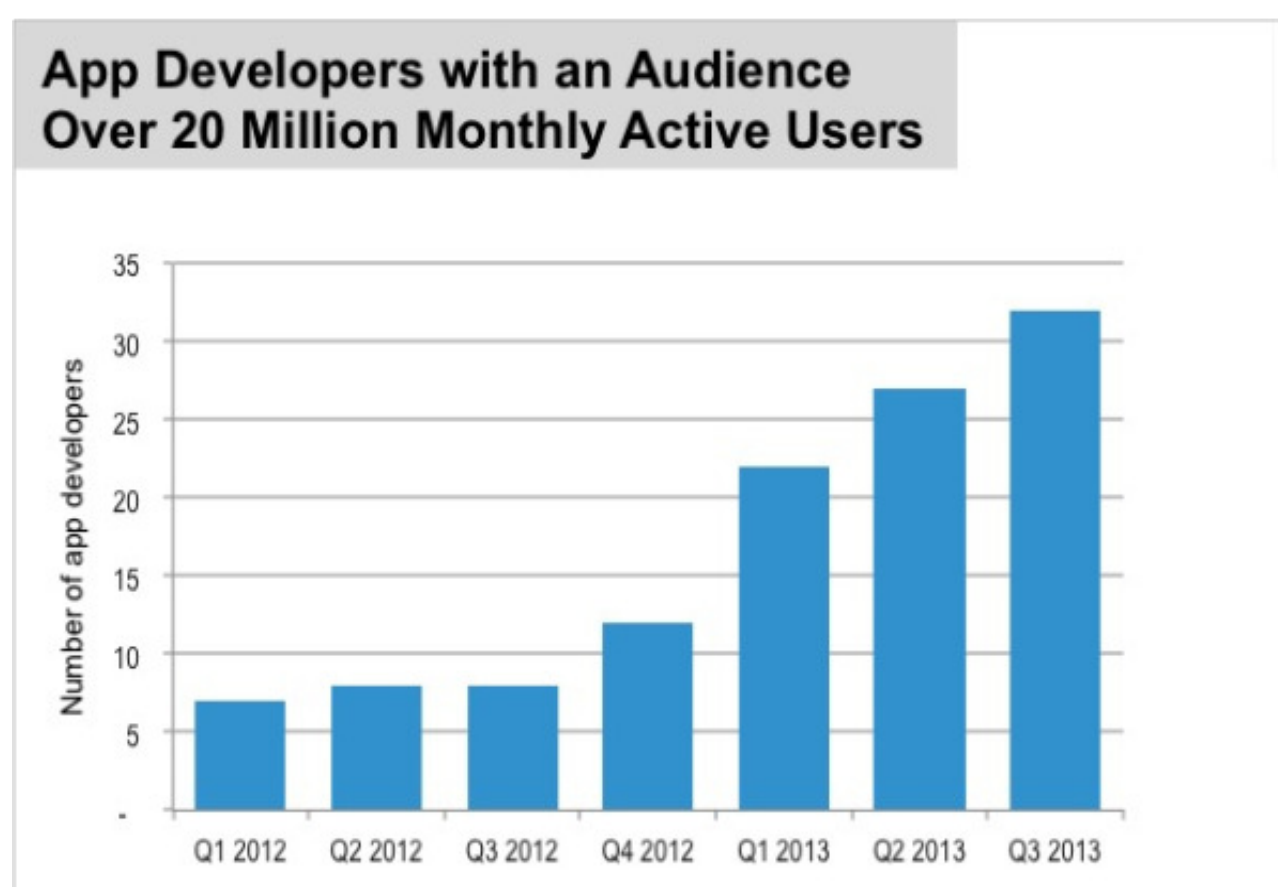

Fig1. Cloud Computing

The above statistics are in conformity with the information collected from apps running on "Flurry Analytics" platform from Q1 2012 to Q3 2013 [8]. The graph in figure (1) below clearly designates that the application starts (as these new apps' appearances are being predicted) have proximately doubled [flurry analytics]. Therefore, there has been a steady growth in the number of apps as presaged by different market researchers

With the incrementing popularity of mobile apps, there has been a consequential magnification in the number of mobile app developers, as well. From the data recorded by flurry analytics, the number of mobile app developers with an audience over one million monthly active users, has gone from 400 to 875 in the same duration, i.e. from Q1 2012 to Q3 2013 as visually perceived in figure (1). The number of mobile app developers with an audience over 20 million active monthly users has additionally experienced an incrementation as observed in figure(2). This elevate in the number of app developers additionally suggests that there are several incipient apps being integrated into the already subsisting ones.

As pictured above most of the statistics only represent the incrementation in the number of apps or the number of users and number of developers every year, but not much effort has been made to record the data on the number of users that are deleting a particular app, the average life of an app, i.e. how long does an average user retains a particular app on his/her smart phone; and how often do the users switch to other apps. Research by Mobilewalla revealed that users eventually expunge 90 percent of all downloaded apps. One erroneous move that vexes or frustrates users – and chances are the app will be expunged. The next few sections in this paper make an effort to verbalize about the factors causing bad apps, which would provide insights on why users are prompted to expunge apps.

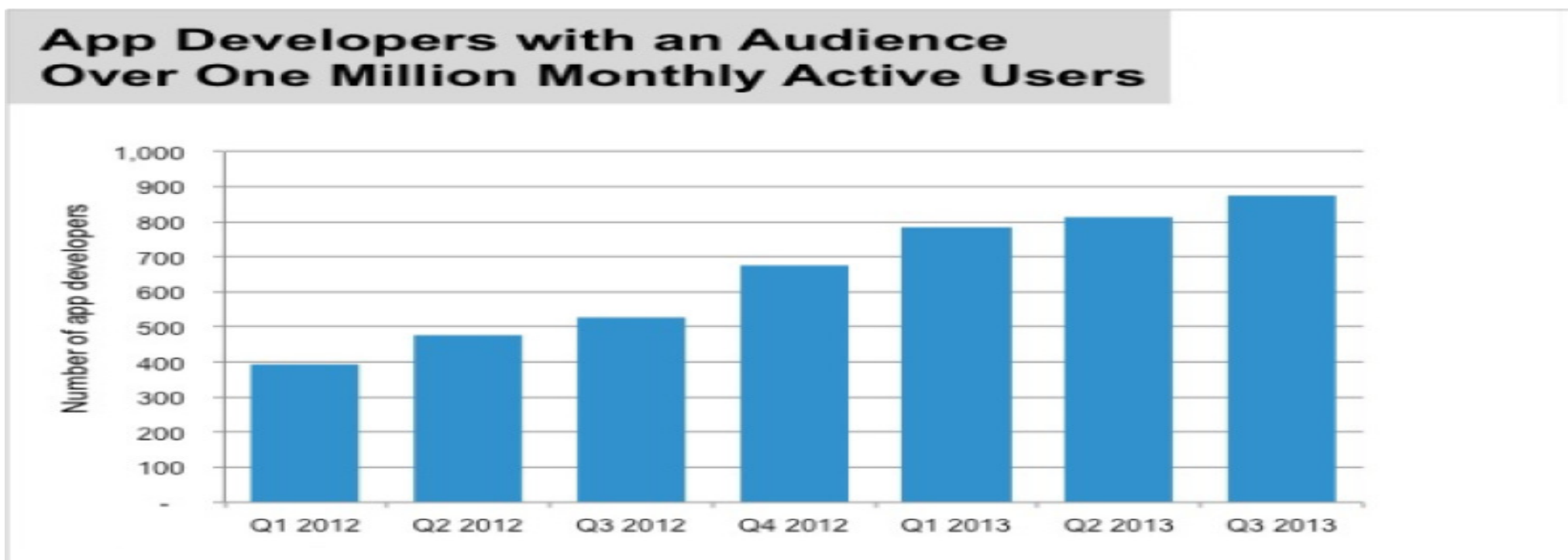

Figure 2: Developers vs Users

# 3. BAD APPS

According to the top negative reviews and statistics [5], “44% verbally express they would expunge a mobile app immediately if app did not perform as expected”. The numbers clearly point out that there are good apps, and lamentable apps in the app market. App users not only uninstall the app, but withal provide negative reviews on the app when customers do not relish the app. With social media and word of mouth being so popular negative reviews spread rapidly, which rigorously affects the reputation of developers and poses a threat to their future releases. Hence it is very critical for the developers to understand the criterion for good apps vs. bad apps and develop accordingly. a) Volume: Many factors contribute towards increasing Volume - storing transaction data, live streaming data and data collected from sensors etc.,

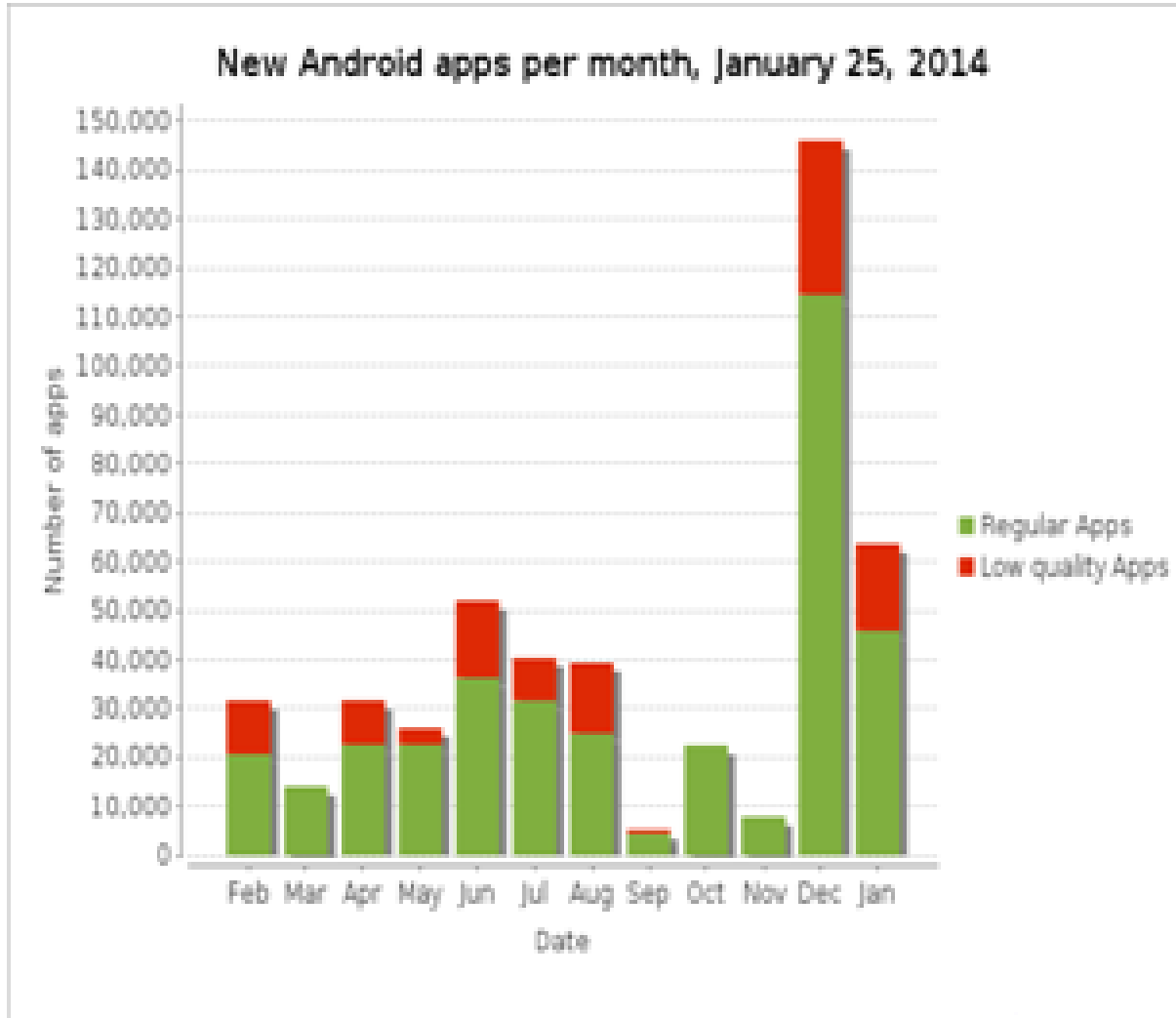

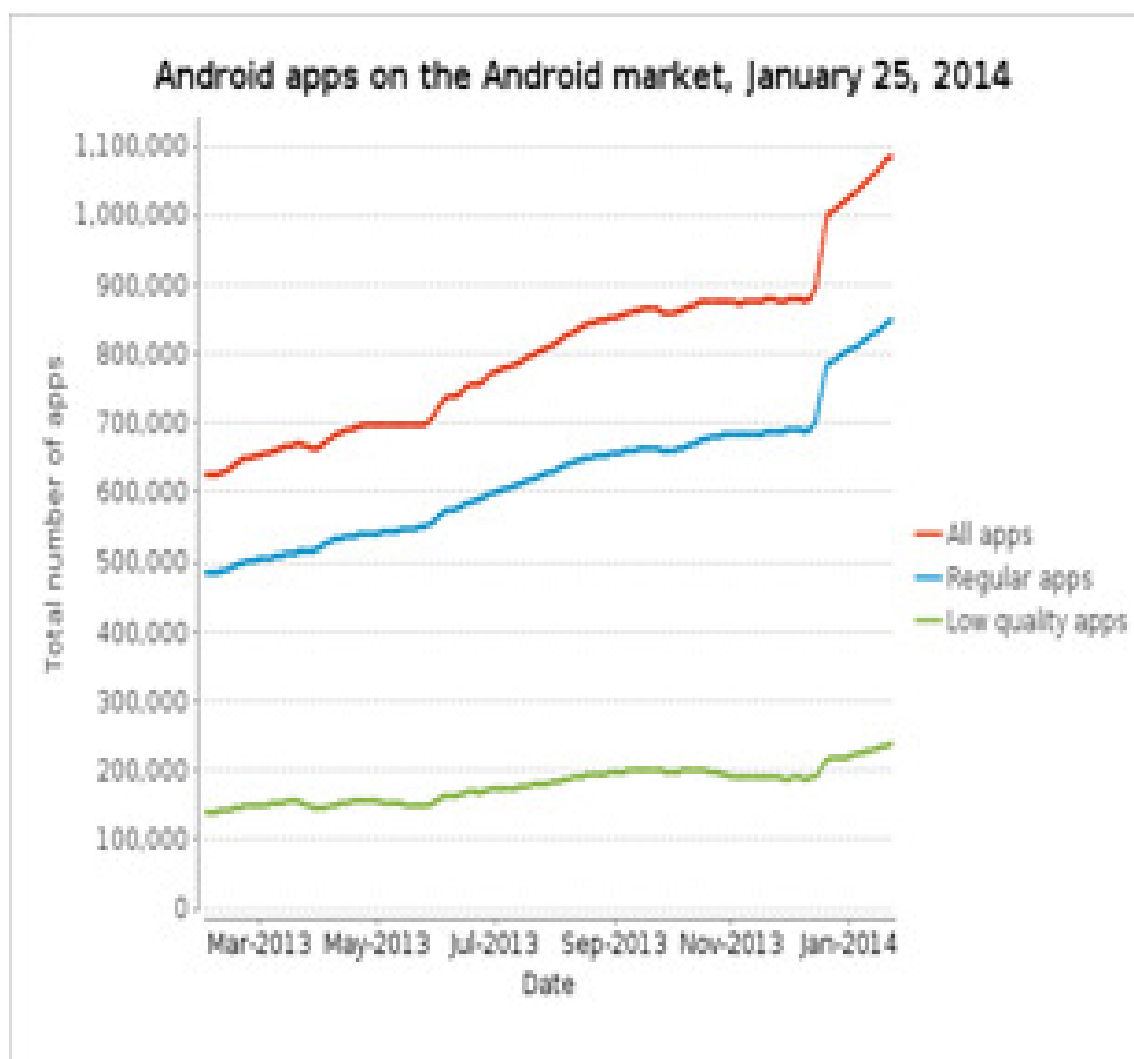

Figure 3: All apps vs low quality apps

Based on the recent information from [7] it is lucidly evident that with the rise in the number of apps (android) ,the number of deplorable/low quality apps are additionally rising as depicted in figure (3). The percentage of low quality apps (22%) as mentioned on [7] is a very consequential number and shows that there is an immediate need to fixate on the quality of apps. With several apps on board and with hundreds of thousands of options, the consumers are persuaded to buy highly polished apps, and so the market is no longer favorable to amateur apps.

*What are bad/low quality apps?*

“Bad” refers - not conforming to standards. An app can be considered to be deplorable if it has a poor design/UI(naïve developers),has lot of clutter, has poor navigation, does not meet the user requirements, does not address a specific issue, has security issues, fails at certain essential times, has downloading issues, is not consistent across various platforms, has compatibility issues, consumes lot of battery power, has a very slow replication function, has very high ad frequency, is not appropriately priced, and has no endeavors made to fine-tune issues/concerns raised by the users. Apps must be updated regularly to keep customers focused and engaged. If an app is made and has never been looked at again, then the app could be counted as a bad app.

# 4 REASONS FOR BAD / FAILING APPS AND SUGGESTIONS TO IMPROVE

According to [2], 22% of mobile users utilize an app only once after downloading it from the app store. This number denotes that there is an immediate need to amend the quality and the functions provided by the app. If the app is being used only once, then the app betokens that it is not engendering or integrating any value to the user. Several factors are influencing apps to fail and to be considered as deplorable. Some of the factors are listed below, and suggestions have been given that would avail in minimizing the number of deplorable/failing apps.

## 4.1 App failure – Role of the developer

Developers should take enough care while building an app. In particular, engendering an app for an organization's brand/product/service requires supplemental attention. Mobile App is one of the paramount marketing implements for any product/service. It might build/eradicate the brand equity and brand adhesion, according to its performance. Additionally, further attention is required when developing critical apps like banking app since it is very arduous to convince customers regarding the security of such apps. Cumbersomely hefty Advertising should be eschewed on apps like banking because, popping up of ads during the course of a transaction might frighten the users regarding their personal banking security.

Developers are the first and foremost ones who are responsible for the failure of an app. There are various causes which include lack of expertise in terms of app development, minimal resources, minimum/no knowledge of user demands and expectations, no knowledge of target audience, and lack of communication between developers of an app. Below mentioned are some suggestions that would aid developers to hold off from making bad apps.

Suggested Solutions :

- Developers should follow the process oriented approach while developing a mobile application. Though the process is sometimes time consuming, it is easy to refer back to the process and rectify the app if any errors are reported [13,14,15,17]. A good design process on-boards the user during the first installation and allows them to personalize their experience, thereby forcing some investment and a more fitting experience on future visits.
- Developers should accumulate input and information from the users through surveys and other designates of research effort developing a mobile app. In other words market orientation is very paramount. This process would avail developers to build apps that would integrate value to the users and will withal eschew redundant apps.
- Developers should possess adequate training and enough practice before beginning to develop apps.

- Developers should use well known and certified tools during development of an app.Superior quality apps can be developed through effective tools.
- Developers should spend sufficient time on testing the apps with respect to security and performance criterion which will play a vital role in the success/failure of an app.
- Developers can use any software management tool to plan their mobile app development activities. Efficient planning will help the developers to meet strict deadlines.
- Developers should integrate user feedback into subsequent versions of the app to remove non-obvious blocks to sustained usage, so developers need to treat user questions and comments like high-valuable unpaid consulting opinions.
- Small things such as push notifications and alerts should be used carefully. They will keep the app top of mind when they are used responsibly to convey relevant content to the user.

- Developers should avoid having complex registration process. Many users return in between the process of downloading the app if the registration process is complex and consumes more time.

## 4.2 App failure-Role of the User

Users/consumers own a significant role in the success/failure of an app [16]. A Mobile App review survey was conducted by [5] which had a sample of over 500 American mobile app users, aged 18 years or older. According to the results of the survey "96% of the American mobile app users say there are frustrations that would lead them to give an app a bad review" [5], including the following:

- Application/system freezes – 76%
- Application/system Crashes – 71%
- Slow responsiveness – 59%
- High battery consumption – 55%
- Considerable amount of ads and promotions – 53%

The survey also has recorded the statistics of the number of users for whom performance matters the most. Without any doubt, the number is 98%, i.e. almost every app user considers performance as his/her main priority. When the users were questioned about the type of apps, for which the performance mattered the most: 74% said banking apps, followed by maps (63%), mobile payments (55%), mobile shopping (49%), games (44%) and social media (41%). Interesting statistics have also been recorded regarding the consequences of poor performance of apps [5]; 44% of the users would delete the app immediately, 38% would delete the app if it would freeze for longer than 30 seconds, 32% would use a negative word of mouth to inform about the bad performance, 21% would post their negative comments on social media (such as facebook, twitter, and blogs), 18% of the sample responded by saying that their patience time is only about 5 seconds, i.e. they would delete the app if it froze even just for 5 seconds and this number is expected to increase drastically in the future. Even though the number is small; 27% of people said they would keep a paid app a little longer in spite of its poor performance, but the damage is already done which is not easily repairable. The users would not prefer to buy any such apps in the future from that particular developer/brand. Some suggestions for the users that would avail ameliorate the performance of apps are provided below.

*Suggested Solutions :*

- Most of the times users just delete an app without even caring to provide a reason that propelled them to delete the app. Users need to provide valuable feedback by describing the concerns/issues that had prompted them to delete an app. The feedback would aid developers in understanding the shortcomings of the app and would also help them in ameliorating the app.
- Instead of just rating the app in terms of numbers or just saying good, bad or average users should take some time to describe their positive/negative experiences with the app
- Though not required, users should endeavor to understand that there could be numerous factors that affect the performance of the app such as processors, recollection,memory and compatibility issues. Users should, not expect superior performance on low end mobile devices, as the quality of the app has nothing to do with the performance. Hence the quality of the apps should not always be inculpated when there could be faults with the devices/network.

## 4.3. App Failure-Role of Technical details

Most Mobile app developers and vendors sometimes fail inefficaciously communicating the technical details of the app to the end users, the result of which is a substantial damage to the prosperity of an app. High end games can be efficiently played only on high end mobile devices, but due to the lack of technical details, users would download and endeavor to play the game on non congruous mobile device configurations. Such incompatibility issues would cause the phone to either freeze or crash or respond very gradually. This results in deplorable reviews from the users and damage is done to the reputation of the app.

*Suggested solutions:*

- All the prerequisites to install an app should be provided in the details of the mobile app which would avail users to download the apps that are suitable and compatible with their devices.Such information would minimize lamentable performances and negative reviews.

## 4.4 App failure-Role of Marketing

With so many developers building new apps each day, it is becoming extremely difficult to acquire/attract, retain and monetize customers [9] and to develop develop brand equity and loyalty. Developers and apps are similar to brands who need to be marketed.App marketing plays a vital role in the success/failure of a mobile app. Marketing and social efforts are required to keep consumers engaged after the app is downloaded to their device.

Insufficient marketing efforts and marketing strategies will lead to decline in return on investments and hence will result in disappointments and frustrations. 70% of developers are frustrated with the current state of app marketing [9]. Below are some of the factors that cause marketing efforts to fail and suggestions are provided in this efficacious marketing.

- Tight Budget: Budget is one of the major issues that developers are concerned. To reach larger audiences, developers need to work with some ad agency providers to promote their apps. However, working with ad agencies requires an investment which is sometimes huge. The graph below by [9] clearly shows that budget is of major concern to the big, medium and small developers. This concern leads to frustration at times and the graph below is in accordance with the same.
  *Suggested Solution*: Ad network providers ad agencies should come with strategies that would minimize the marketing the costs while maximizing the revenues/ROI (return on investments) so that developer's can relax on the distribution aspects and concentrate more on developing superior apps
- Lack of Trust in the ad network providers: Most of the developer's today are disappointed with the lack of clarity given by ad network providers, and hence are finding it difficult to trust them [9].It is very expensive to advertise an app through agencies and so trust plays a major role while making such investments. Many developers are concerned that the ad network providers are not being honest about their revenue claims with 71% of developers expressing the view that eCPM (effective Cost Per Thousand Impressions) was exaggerated, i.e. eCPM's were used to lure them in but these would often only last for a short period before failing [9].

  *Suggested Solution:* Ad network providers must be more transparent with their offerings, planning, reporting and measurement to gain the trust of the developers for whom ROI and marketing investments are a huge priority. The graphs below indicate the essence of trust in the minds of developers.

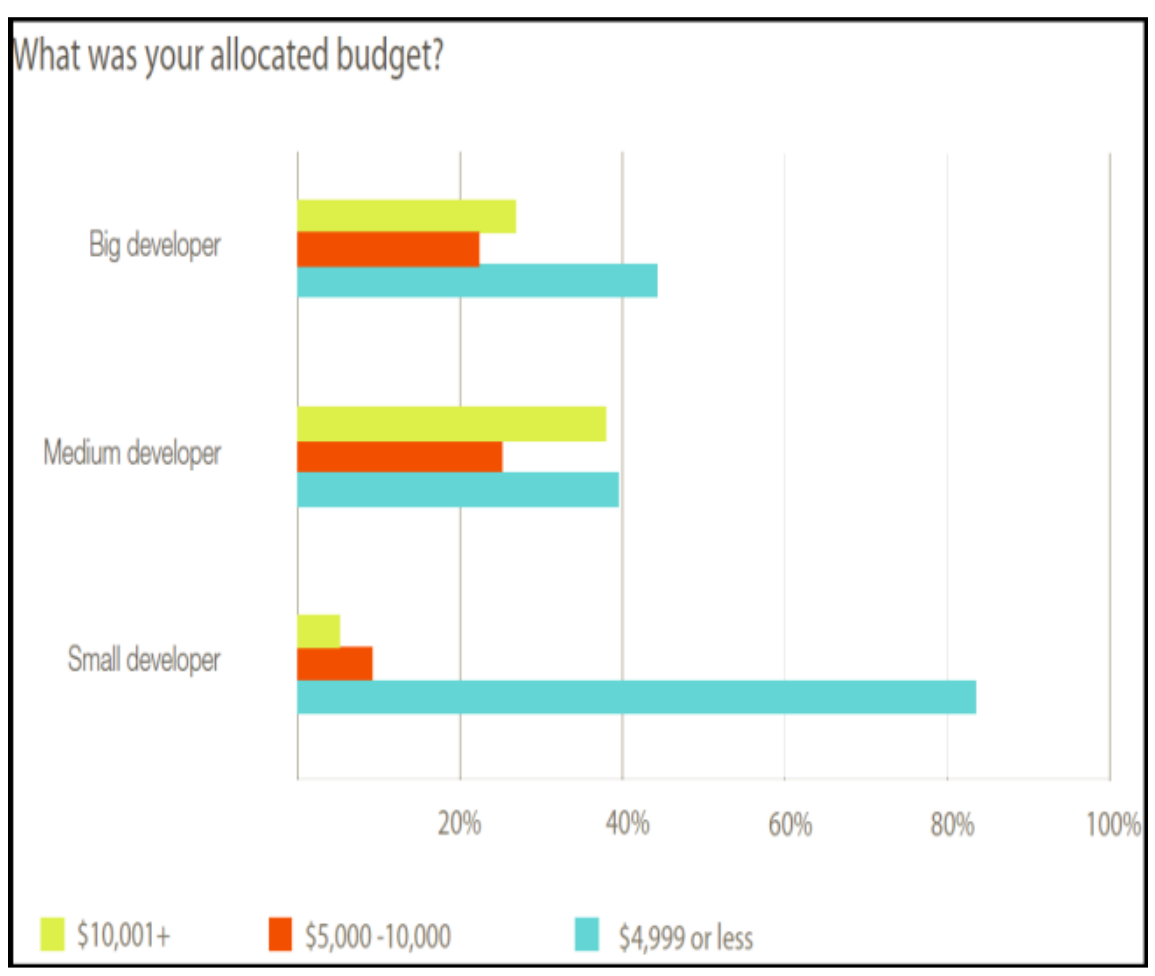

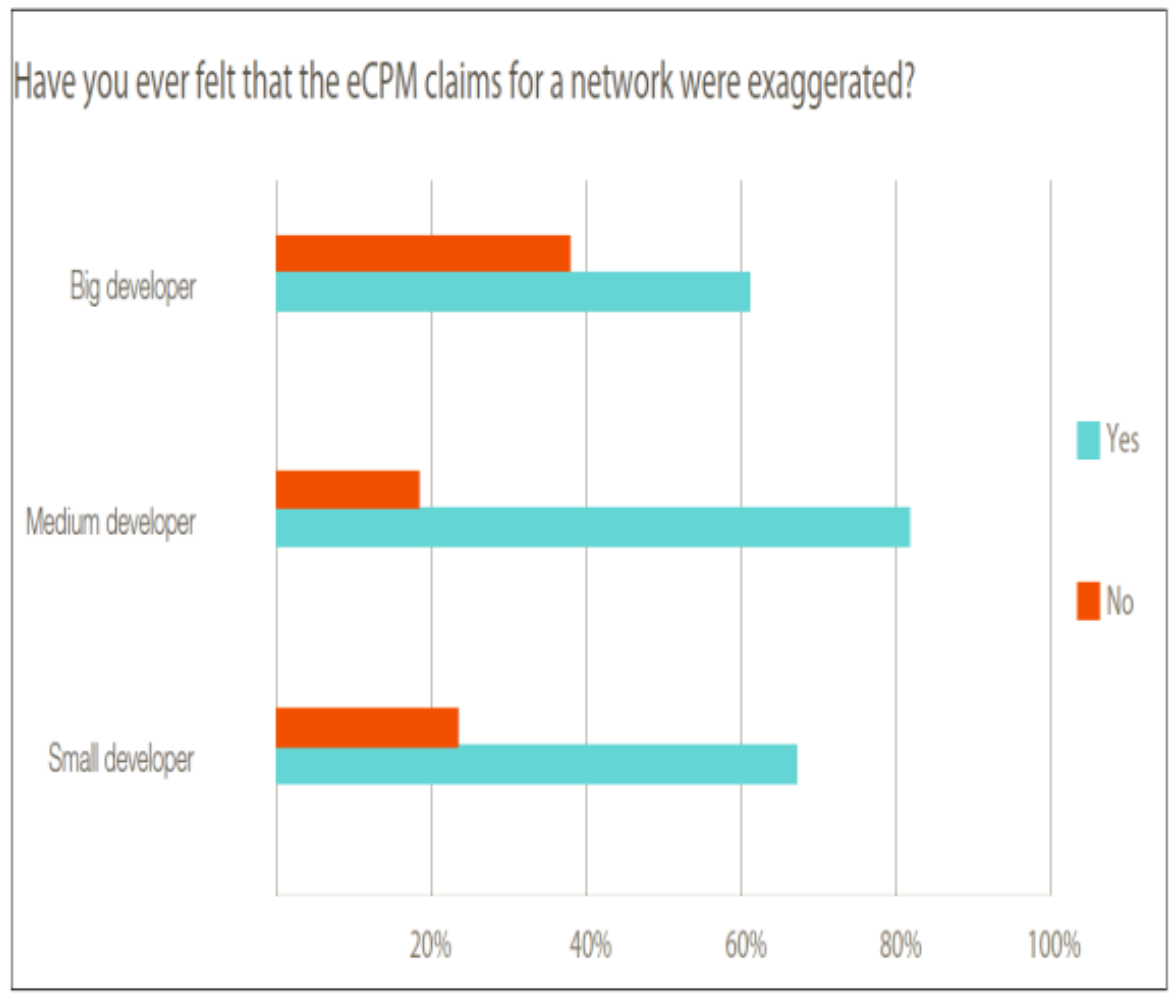
Have you ever felt that the eCPM claims for a network were exaggerated?
Big developer
Medium developer
Small developer
Yes
No
20%
40%
60%
80%
100%

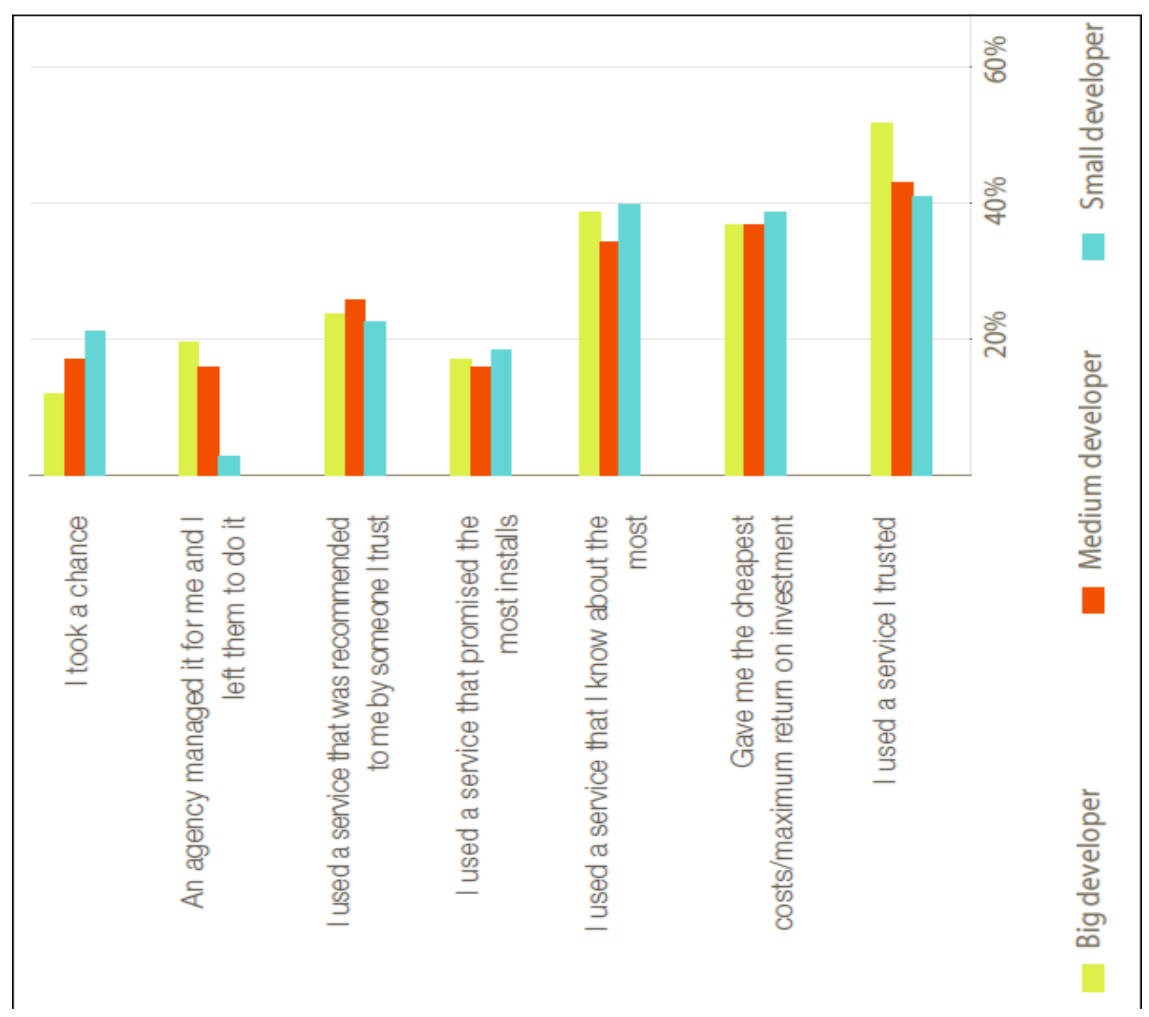
I took a chance
An agency managed it for me and I left them to do it
I used a service that was recommended to me by someone I trust
I used a service that promised the most installs
I used a service that I know about the most
Gave me the cheapest costs/maximum return on investment
I used a service I trusted
20%
40%
60%
Big developer
Medium developer
Small developer

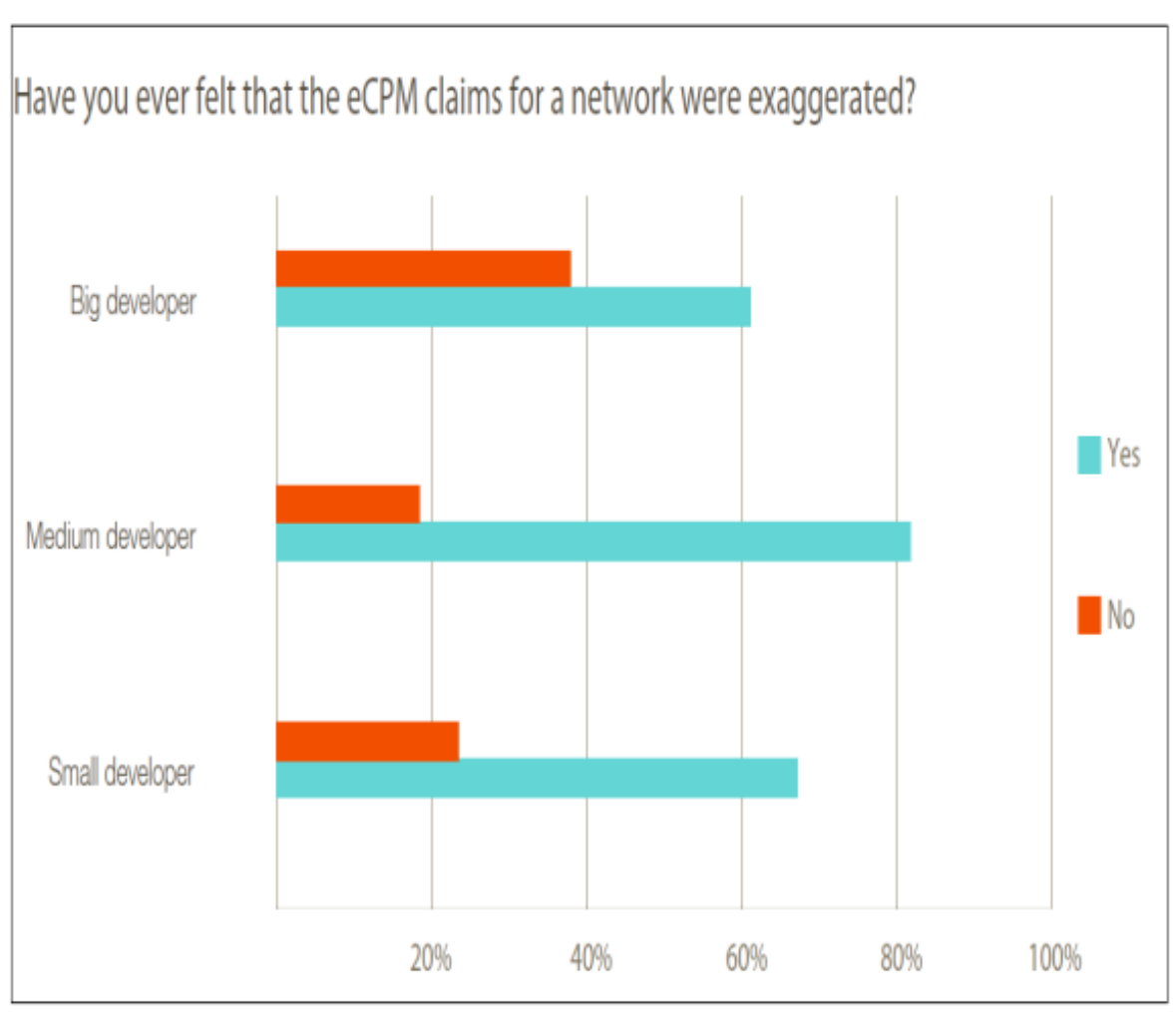
Have you ever felt that the eCPM claims for a network were exaggerated?
Big developer
Medium developer
Small developer
Yes
No
20%
40%
60%
80%
100%

Figure 4: Developer's Concerns regarding marketing of an app

- Physical cues that grab attention: One of the biggest challenges in mobile is the application discovery and awareness. Having an eye pleasing/eye catching visual elements, i.e. an app icon, logo, screenshots is an effective marketing tool. So app icon is the first visual cue to grab attention while users are in a sea of apps [6]. "First impression is the best impression". User reviews should be encouraged within the app,because along with the other physical cues, reviews are the cues that users look for before installing an app.

- Use of Social Media/Viral Marketing: Options that would help users to spread the word through email/social media (like tell a friend, share via Facebook, post on Twitter) and also options that would prompt the users to provide their ratings about the app should be included in an app.WOM and WOSM(word of social media) are free marketing tools that can be used to gain popularity for the app("It may lead to quick negative word of mouth if the quality of the app is poor!!) [6].

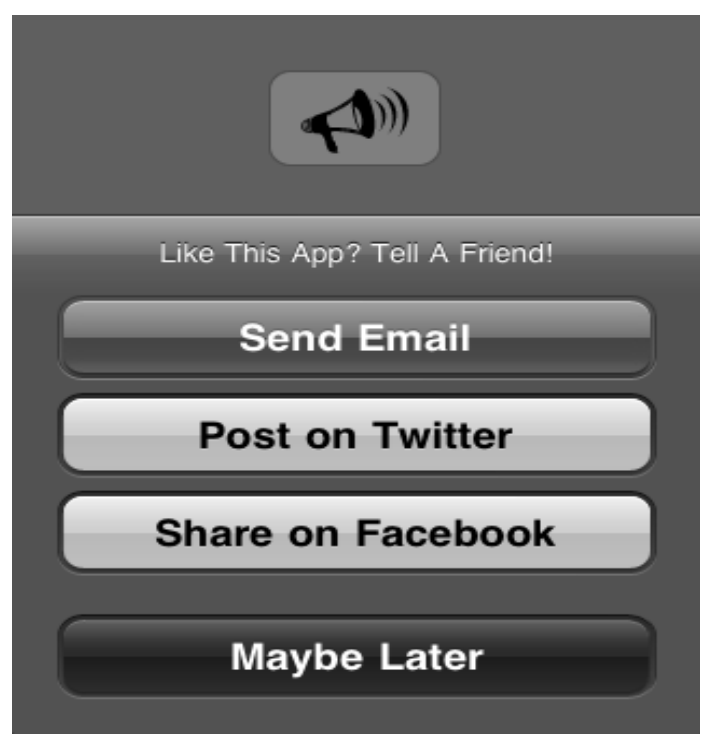

Figure 5: Use of Social media to promote the app

- Use of effective target audiences: Choose target audiences (such as college students) who would help in rapid spreading of information about the app and opinion leaders who would persuade/influence others(friends, classmates, peers) to use the app.
- Pricing: Freemium vs Feemium, "Freemium"(free basic features,additional fee for premium features) is the best way to persuade customers to use your app."Feemiums"( fee to be paid even for basic features) is not an optimal way of introducing an app. If Feemiums need to be used, at least a free trial version should be provided for a couple of days so that the user can decide whether or not to spend on the app. Freemiums do help in getting good revenues when users like the app and are interested in spending a little to continue;using;the;app.
- Use "popular blogs" of friends/family who would advertise your app on their website for minimal cost (or sometimes for free) rather than using heavily priced advertising agencies to market your app.
- Marketing through clients: Working for clients (third party companies) to develop apps for their organization/product/accommodation will avail gain free publicity/marketing support. Providing the clients with superior apps will not only avail in reinforcing the relationship with the client, but will withal avail in income opportunities. [6]. It is a free medium to establish oneself and one's app as a brand.

The above mentioned list is not exhaustive.The ad network providers and developers should work collaboratively in marketing of apps so that users can access and experience superior high quality apps.

# 5. ROLE OF APP DEVELOPMENT LIFE CYCLE

Most of the studies and articles [22,23] verbalize about the statistics of mobile apps in terms of revenues and number of apps being developed, but very few or none of them talk about good quality *vs* low quality apps nor the factors that cause low quality apps nor the solutions to surmount those shortcomings.

What are the reasons for low quality apps from SDLC point of view?

- The first and foremost reason is that the app developers are not conforming to the development life cycle phases. Most of the app developers start developing the app without accumulating requirements and without having a design
- Lack of training and experience on the app development SDKs.
- Not enough testing is done. App developers are more fixated on functional aspects of the app and hence they sometimes ignore security and performance testing, which are the key components of any app.
- Poor maintenance.

Why do we need a software development life cycle and what happens if we do not use systematic approaches while developing the software product? The result is lower quality software products. A mobile application is nothing but a software product with a different level of complexity. One can apply same conventional methods/methodologies (such as waterfall, iteration, agile, and scrum) [18,19,21] along with different mobile app techniques and tools to design, develop, test and deploy a mobile application.

## 5.1 Related Technologies

### 5.1.1 Unified Modeling Language (UML)

Grady Booch, Ivar Jacobson and James Rumbaugh developed the "Unified Modeling Language (UML)" [25] at Rational Software in the 1990s. Unified Modeling Language is an object modeling and specification language used in software engineering. In the field of software engineering, UML is considered as a standardized general purpose modeling language according to ISO/IEC 19501:2005. The main advantage of UML is that, it creates visual models of system/ object oriented software intensive systems. UML contains a large set of graphical representation techniques. Unified Modeling Language can be used in combination with various modeling components such as object modeling, data modeling, component modeling and business modeling.

### 5.1.2 Mobile Unified Modeling Language (M-UML)

For modeling mobile agent-based software system, UML cannot be used as it does not possess the mobility requirements. The mobile agents carry an executable code and data within themselves. An extension to UML has been defined by Kassem Saleh and Christo El-Morr for mobile agent systems, which is known as Mobile Unified Modeling Language (M-UML) [26]. Mobile agents became more feasible due to the advancements in remote evaluation, process mitigation, distributed object computing and mobility.

For a mobile agent-based system, M-UML can be well described by going through each of the UML diagrams and explaining the modifications and extensions that are required for describing the mobility aspects. In M-UML, the authors introduced two major extensions to the existing UML. The first one represents Mobile process/service with the letter "M" at the top corner of the diagram and the other one is, represents the remote process/service with letter "R" at the top corner of the specified diagram(could be a Use case/Sequence/Class diagram).

## 5.2 Case Study

The social networking application is one of the most popular and top ranked application on mobile devices [22,23]. Social networking is a network made of people and organizations who communicate socially/professionally.Facebook [28] application is one of the popularly used networking site for sharing social events, messages, multimedia content like videos and photos. Google+ [28] is another social networking application like Facebook which is associated with Google services like mail, YouTube, etc.,A social networking application is not just intended for social life, but likewise assists to amend the professional views of life with applications like LinkedIn [28], a professional platform where individuals with dissimilar and similar skill sets can join together and raise their quality of work. Twitter [28] is another social networking media popularly known for its micro blogging services.

All the social networking applications share some basic functionalities which include signing into a social networking app, addition/deletion of a known/ unknown profile, sending/receiving a message, sharing text information, sharing multimedia information and location based services [27,28].

*What is Generic Social Networking (GSN) App?*

A GSN application can be defined as the one that holds the basic characteristics of any social networking app. GSN app can be employed as a basic template for developing any social networking app in the future, i.e. for any app to be a social networking app, it requires to at least have all the features of GSN. This paper will talk about the development of GSN app, using the traditional object oriented software development life cycle approach. The approach includes the following phases: requirements gathering, design, development, testing and maintenance. This process would help improve the quality of the app. UML, M-UML and development tools will be used to explain the mobile app development life cycle.

### 5.2.1 Mobile App Requirements and specification

Requirements phase is the first and foremost phase of mobile application development life cycle, which deals with functional and non-functional aspects of the application [18]. A functional requirement describes thefunctionalities/services of an application.Functional requirements of any mobile application are similar to that of a traditional software application in the context of requirements gathering, and hence the question here is how can one distinguish the requirements of mobile app from the traditional software application. In order to resolve this issue, M-UML language introduced by Kassem Saleh and Christo El-Morr will be used in order to represent the features of mobile application.

Mobile app requirements can be specified /modeled with the help of M-UML. Most of the apps interact with remote web server/ web services which can be specified /modeled with the help of UML. In figure (6), the high level mobile app use cases and actors are shown. One can easily distinguish between a mobile actor and other actors. Mobile app use cases are represented by the letter 'M' on the top corner of the use case. A mobile app use case describes the functional requirement of the application and should have at least one mobile actor interacting with the use case. The other features like extending the use cases, reusing the existing use cases work in a similar fashion to the traditional use case modeling.

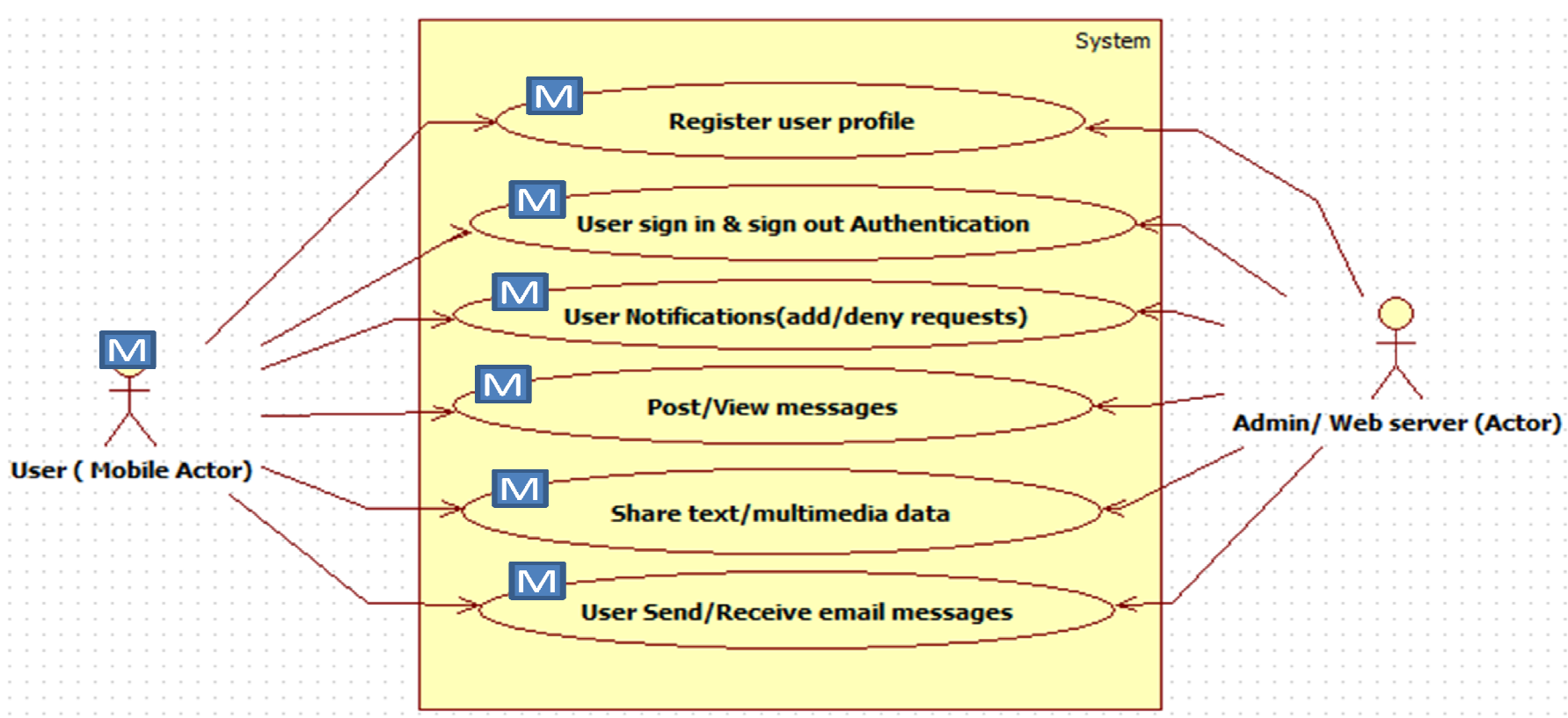

Figure 6. GSN mobile app use case modeling

In figure (6) high level GSN mobile app use case system is shown, which consists of various mobile use cases such as register user profile, user authentications (such as sign in, sign out), user notifications (add/deny user requests), send/receive mails, and share messaging/multimedia data. Most of these mobile use cases are generic to any social networking apps, and hence it is named as a GSN app. These mobile use cases can be combined or further divided into different use cases. The primary motive of designing this use case is to get a clear understanding of the system in the first place. Mobile use cases can be described as shown in figure (7).

### 5.2.2 Mobile App Design Phase

Mobile App design phase, is a critical phase and is carried out in two designs:Static design and Dynamic design. Static design describes the static nature of the system in the form of class diagrams where as dynamic nature is described with the help of sequence and collaboration diagrams.

#### 5.2.2.1 Mobile app static design

Generic social networking app static design can be explained as in figure (8), which contains a set of classes, and if any class uses/initiates any of the mobile objects then it becomes a mobile class. A mobile class is represented with the letter "M" at the top corner of the class representation. Each mobile class is further divided into attributes and operations like any other object oriented class concept. A mobile class may interact with remote class/remote object which is denoted with the letter "R" at the corner of the class presentation, to differentiate between a mobile class and a remote class.

| **Mobile Use Case Name** | User notifications |
|---|---|
| **Summary** | The user should be notified of various kinds of activities on his/her account |
| **Mobile Actor** | Mobile app User |
| **Precondition** | User needs to sign into a mobile app |
| **Description:** | 1. Any kind of activity in the user profile is logged<br>2. There are two types of activities in the system one from the user side, and the other from the people in a user's network.<br>3. Whenever there is an event on the user's network, it should be notified to the user immediately. This includes messages, news, and invitations.<br>4. The user should be able to read and/or delete the message |
| **Alternatives** | 1. Some notifications are accessible by user as people in the user's network can block/personalize their notifications. |
| **Post condition** | User able to access notifications |

Figure 7. User notifications, mobile use case description

User profile, User account, User notifications, Data handler and User mail messages are some of the important classes that are identified to implement the app. Figure (4) shows association between classes and kind of participation (multiplicity) in the relationship. The Data handler class has a hierarchy (inheritance) relationship.

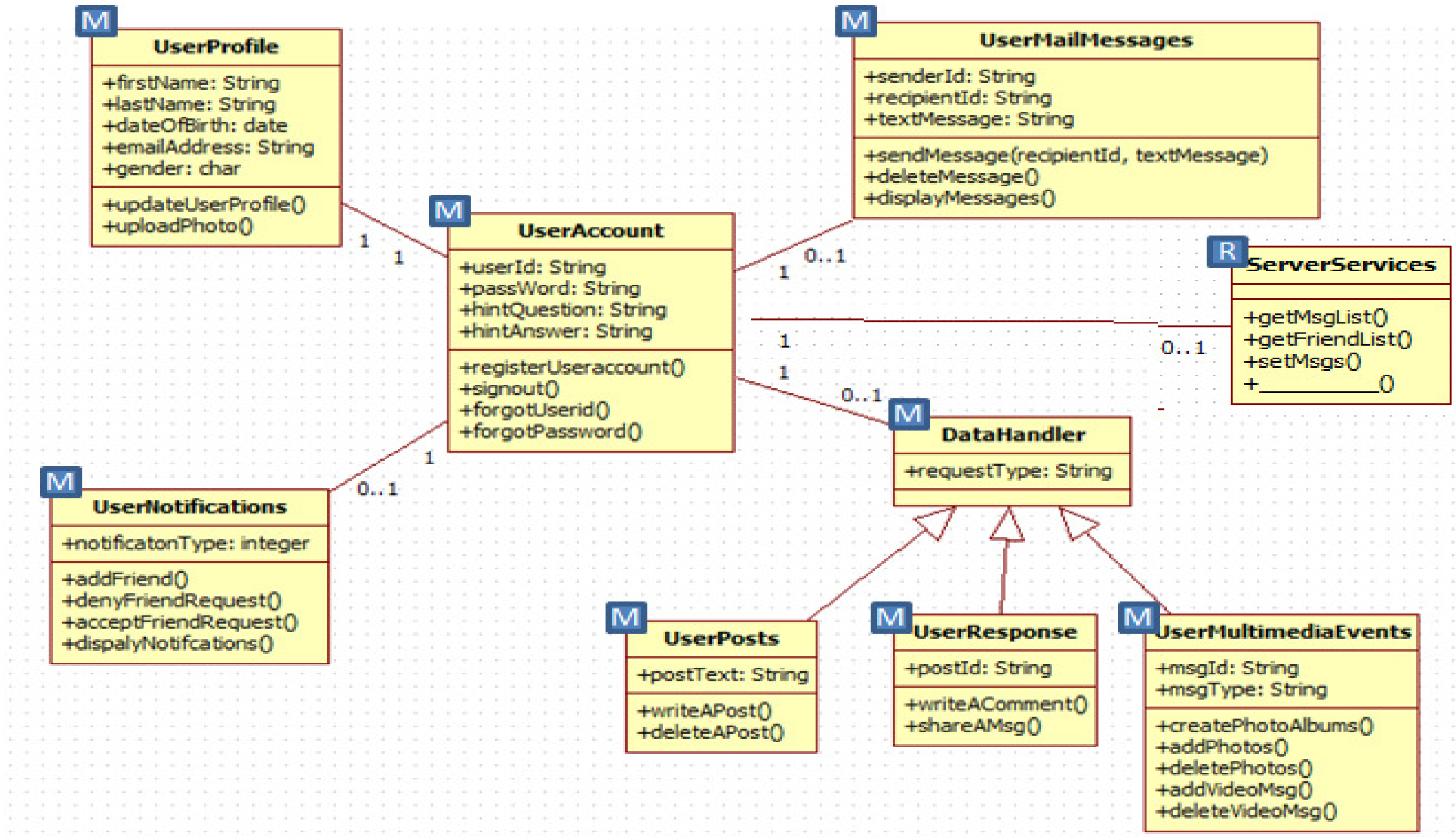

Figure 8: GSN mobile app class diagram

### 5.2.3 Mobile app dynamic design

The dynamic nature of the mobile app can be described by using either sequence diagrams or collaboration diagrams. Time based interactions are represented using sequence diagrams and association between objects are represented using collaboration diagrams. Sequence diagram for the mobile use case (post message) is represented as shown in figure (9). When a mobile app user signs into the app using his/her credentials, the remote server validates the mobile user credentials, and then returns a valid/invalid user message. If the user is a valid user then he/she can write messages or post, and then the message/post is sent to the server and displayed accordingly on the user's profile.

Fig (9), explains the collaboration diagram for user notifications use case.When a mobile user selects the notification, mobile class sends the requests to the server. Depending on the server response, notifications are displayed to mobile users.

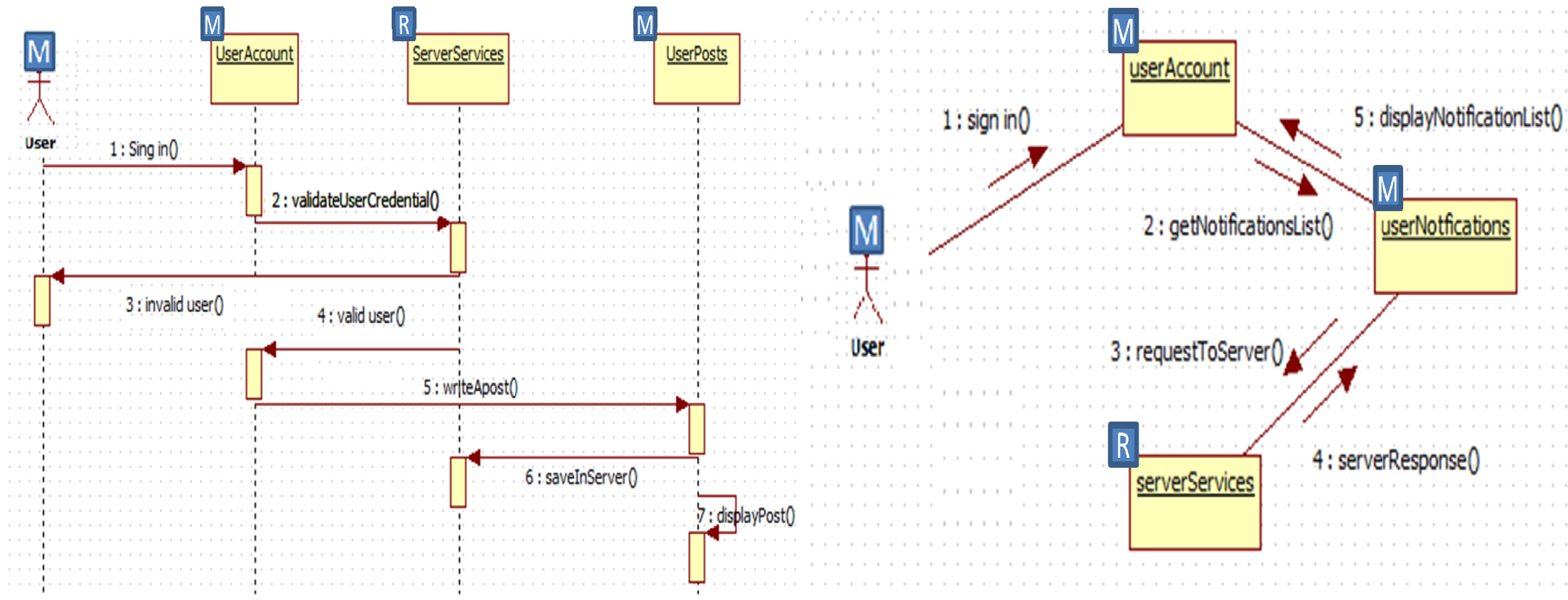

Figure 9: Post message sequence diagram & User notifications collaboration diagram

### 5.3 Mobile App Development Phase

Developing a web application/traditional software application involves certain kinds of actions which include finding the right/appropriate technology (programming language) and choosing a specific platform (such as Windows, Linux, Mac). Development of mobile app is similar to a traditional software application development with some challenges and complexities involved due to software and hardware constraints [20].As mentioned earlier, the mobile application market is evolving tremendously and hence academic researchers and companies are creating new platforms and programming tools for the mobile app development. Among the different platforms that exist: Android, IOS and windows are most popular and are widely used [28]. Android supports the Java programming language, IOS supports object c and pascal, and windows support c, c++ and visual basic languages [28,29]. Each program has its own cons and pros regarding technology and marketing prospects.

GSN is developed (demo version, only for research) with the help of android (SDK) [29], which is an open source and has rich set of supporting tools and plug-ins that makes the app more flexible and robust. The primary motive of the development of GSN app is to show the interaction between the mobile classes and remote classes that supports our Mobile design in section 3.2.

### 5.4 Mobile App Testing Phase

Mobile app testing plays a vital role in determining the quality and performance of the app. In order to deliver superior quality apps, efficient techniques and testing tools have to be applied. Each platform provides a certain set of testing tools by default. Apart from the default tools, there are plenty of tools available in the market for testing mobile apps.Analysis reports from different market statistics [22,23] suggest the following four important testing aspects to avoid low quality apps. Functional testing (black box testing), Code testing (white box testing), security testing [30,33], and, performance testing [31]. Mobile application testing can be carried out, either by using emulator tools or by using a real mobile device (phone/tablet)

For GSN app, we have derived 20 functional test cases based on mobile use cases. The test cases have been tested/run using the emulator which is a default functional testing tool for Android SDK [29]. The code testing was done with the JUnit tool which is white box testing. Security and performance testing was done with the help of android SDK tool kit. There are several tools available in the market that avail to amend the quality of the mobile apps in every phase of testing [24].

### 5.5 Mobile App Maintenance Phase

After the development and testing of the mobile app, vendors release the mobile app into the market for mobile users. There is a high probability that users might face few issues that were not identified during the mobile app testing phase. This is a very mundane scenario that transpires in most software product releases, for various reasons, including device compatibility, software and hardware constraints, and, network problems.Maintenance phase deals with fixing the issues that were fronted by the mobile users and also involves developing/releasing new features, which can be implemented using mobile application development life cycle. Integrating new requirements in the mobile use cases, projecting, developing and finally testing improves the overall caliber of the app when a systematic procedure is being implemented.

## 6. RESEARCH REFLECTIONS

The primary motive of this research is to show that the methodological aspect of software engineering can be successfully applied to mobile applications, with the help of the software development life cycle. Following are the observations of this research:

- Mobile app is not a tiny/simple piece of software anymore.
- Apps are continually growing in number and complexity.
- Process oriented approaches and techniques are required to handle mobile application development [32,33].
- Traditional software approaches and methods can still be applied to mobile application development.
- Object oriented concepts and methodologies can be applied to mobile application development.
- Superior quality mobile applications can be developed with the help of process oriented methods.

## 7. SUMMARY AND FUTURE WORK

With the App Store is such a phenomenal success, mobile apps have become an essential part of the user's daily life. To obtain prosperity in the current mobile world, merely developing an app is not just enough. Thus, the goal of the developer is to keep an open mind and embrace innovative procedures that can avail amend the life cycle (gathering requirements, design, development, testing, and maintenance) of an app while enhancing user satisfaction. New technologies are emerging every day, and hence developers should always be well apprised about current trends, requirements, and events in the mobile technology field. There will always be incipient avenues that will arise in the future for developing new and innovative apps. With such new and promising opportunities arising every day for mobile app developers, the future assures to be an enticing journey [22]. Network with the integration of the wind generators and the SVC devices . Apart from the development of mobile application, we have made a few observations (research reflections) that surface lot of scope for the future areas of research on mobile application development